\documentclass[twocolumn]{article}
\usepackage{jeffe,myconcrete,euler,epsfig}
\usepackage[nocopyright]{jeffeproc}
\usepackage[tilde,same]{url}

\newtheorem{lemma}{Lemma}[section]
\newtheorem{theorem}[lemma]{Theorem}
\newtheorem{corollary}[lemma]{Corollary}
\newtheorem{conjecture}[lemma]{Conjecture}

\def\sd{\operatorname{sd}}
\urldef{\paperurl}\url{http://www.cs.uiuc.edu/~jeffe/pubs/spread.html}

\begin{document}

\title{\bf Nice Point Sets Can Have Nasty Delaunay Triangulations%
	\thanks{Portions of this work were done while
	the author was visiting INRIA, Sophia-Antipolis, with the
	support of a UIUC/CNRS/INRIA travel grant.  This research was
	also partially supported by a Sloan Fellowship and by NSF
	\hbox{CAREER} grant CCR-0093348.  See \paperurl\ for the most
	recent version of this paper.}}

\author{Jeff Erickson\\[1ex]
	\normalsize
	\begin{tabular}{c}
	University of Illinois, Urbana-Champaign\\
	jeffe@cs.uiuc.edu\\
	\url{http://www.cs.uiuc.edu/~jeffe}
	\end{tabular}
}

\date{}

\maketitle

\begin{abstract}
We consider the complexity of Delaunay triangulations of sets of
points in $\Real^3$ under certain practical geometric constraints.
The \emph{spread} of a set of points is the ratio between the longest
and shortest pairwise distances.  We show that in the worst case, the
Delaunay triangulation of $n$ points in~$\Real^3$ with spread $\Delta$
has complexity $\Omega(\min\set{\Delta^3, n\Delta, n^2})$ and
$O(\min\set{\Delta^4, n^2})$.  For the case $\Delta =
\Theta(\sqrt{n})$, our lower bound construction consists of a uniform
sample of a smooth convex surface with bounded curvature.  We also
construct a family of smooth connected surfaces such that the Delaunay
triangulation of any good point sample has near-quadratic complexity.
\end{abstract}

\section{Introduction}

Delaunay triangulations and Voronoi diagrams are used as a fundamental
tool in several geometric application areas, including finite-element
mesh generation
\cite{ceft-se-99,elmsttuw-scus-00,lt-gsftd-01,s-tmgdr-98}, deformable
surface modeling \cite{cdes-dst-01}, and surface reconstruction
\cite{ab-srvf-99,abk-nvbsr-98,acdl-sahsr-00,ack-pcubm-,bc-ssrnn-00,
hs-vbihc-00}.  Many algorithms in these application domains begin by
constructing the Delaunay triangulation of a set of $n$ points in
$\Real^3$.  Delaunay triangulations can have complexity $\Omega(n^2)$
in the worst case, and as a result, all these algorithms have
worst-case running time $\Omega(n^2)$.  However, this behavior is
almost never observed in practice except for highly-contrived inputs.
For all practical purposes, three-dimensional Delaunay triangulations
appear to have linear complexity.

One way to explain this frustrating discrepancy between theoretical
and practical behavior would be to identify geometric constraints that
are satisfied by real-world input and analyze Delaunay triangulations
under those constraints.  These constraints would be similar to the
\emph{realistic input models} such as fatness or simple cover
complexity, which many authors have used to develop geometric
algorithms with good practical performance
\cite{bksv-rimga-97,v-ffrim-97}.  Unlike these works, however, our
(immediate) goal is not to develop new algorithms, but rather to
formally explain the good practical performance of existing code.

Dwyer \cite{d-hdvdl-91,d-enkfv-93} showed that if a set of points is
generated uniformly at random from the unit ball, its Delaunay
triangulation has linear expected complexity.  Golin and Na
\cite{gn-ac3dv-00} recently derived a similar result for random points
on the surface of a three-dimensional convex polytope.  Although these
results are encouraging, they are unsatisfying as an explanation of
practical behavior.  Real-world surface data generated by laser range
finders, digital cameras, tomographic scanners, and similar input
devices is often highly structured.

This paper considers the complexity of Delaunay triangulations under
two types of practical geometric constraints.  First, in
Section~\ref{S:spread}, we consider the worst-case Delaunay complexity
as a function of both the number of points and the \emph{spread}---the
ratio between its diameter and the distance between its closest pair.
For any $n$ and $\Delta$, we construct a set of $n$ points with
spread~$\Delta$ whose Delaunay triangulation has complexity
$\Omega(\min\set{\Delta^3, n\Delta, n^2})$.  When $\Delta =
\Theta(\sqrt{n})$, our lower bound construction consists of a
grid-like sample of a right circular cylinder with constant height and
radius.  We also show that the worst-case complexity of a Delaunay
triangulation is $O(\min\set{\Delta^4, n^2})$.

An important application of Delaunay triangulations that has received
a lot of attention recently is surface reconstruction---given a set of
points from a smooth surface $\Sigma$, reconstruct an approximation
of~$\Sigma$.  Several algorithms provably reconstruct surfaces if the
input points satisfy certain sampling conditions
\cite{acdl-sahsr-00,ack-pcubm-,bc-ssrnn-00,hs-vbihc-00}.  In
Section~\ref{S:surface}, we consider the complexity of Delaunay
triangulations of good samples of smooth surfaces.  Not surprisingly,
oversampling almost any surface can produce a point set whose Delaunay
triangulation has quadratic complexity.  We show that even surface
data with \emph{no} oversampling can have quadratic Delaunay
triangulations and that there are smooth surfaces where \emph{every}
good sample has near-quadratic Delaunay complexity.  We also derive
similar results for randomly distributed points on non-convex smooth
surfaces.

We will analyze the complexity of three-dimensional Delaunay
triangulations by counting the number of edges.  Two points are joined
by an edge in the Delaunay triangulation of a set~$S$ if and only if
they lie on a sphere with no points of~$S$ in its interior.  Since
every vertex figure is a planar graph, Euler's formula implies that a
Delaunay triangulation with $n$ vertices and $e$ edges has at most
$2e-2n$ triangles and $e-n$ tetrahedra.

In the interest of saving space, several straightforward but tedious
calculations are omitted from this extended abstract.

\section{Sublinear Spread}
\label{S:spread}

We define the \emph{spread} $\Delta$ of a set of points (also called
the \emph{distance ratio} \cite{c-nnqms-99}) as the ratio between the
longest and shortest pairwise distances.  In this section, we derive
upper and lower bounds on the worst-case complexity of the Delaunay
triangulation of a point set in $\Real^3$, as a function of both the
number of points and the spread.

If the spread takes its minimum value $\Theta(n^{1/3})$, the points
are packed into a tight lattice, and the Delaunay triangulation has
only linear complexity.  On the other hand, all known examples of
point sets with quadratic-complexity Delaunay triangulations have
spread $\Omega(n)$.  Thus, it is natural to ask how the worst-case
complexity of the Delaunay triangulation changes as the spread varies
between these two extremes.  The spread of a set of points is loosely
related to its dimensionality.  If a set uniformly covers a bounded
region of space, a surface of bounded curvature, or a curve of bounded
curvature, its spread is respectively $\Theta(n^{1/3})$,
$\Theta(n^{1/2})$, or $\Theta(n)$.  The case of surface data is
particularly interesting in light of numerous algorithms that
reconstruct surfaces using a subcomplex of the Delaunay triangulation.
We will discuss surface reconstruction in more detail in the next
section.

\subsection{Lower Bounds}
\label{SS:lower}

\begin{figure}[htb]
\centerline{\epsfig{file=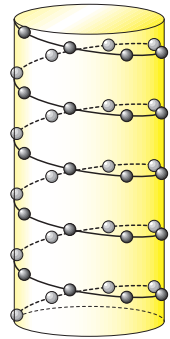,height=1.75in}}
\caption{A set of $n$ points whose Delaunay triangulation has 
complexity $\Omega(n^{3/2})$}
\label{Fig:spiral}
\end{figure}

The crucial special case of our lower bound construction is $\Delta =
\Theta(\sqrt{n})$.  For any positive integer~$x$, let $[x]$ denote the
set $\set{1, 2, \dots, x}$.  Our construction consists of $n$ evenly
spaced points on a helical space curve:
\[
	S_{\!\sqrt{n}}
	= \Setbar{\Paren{\frac{t}{n},\,
			\cos\frac{t}{\sqrt{n}},\,
			\sin\frac{t}{\sqrt{n}}}}
		 {t \in [n]}.
\]
See Figure~\ref{Fig:spiral}.  As we show below, the Delaunay
triangulation of $S_{\!\sqrt{n}}$ has complexity $\Omega(n^{3/2})$.
Note that $S$ is a grid-like uniform $\e$-sample of a right circular
cylinder, where $\e = \Theta(\sqrt{1/n})$.  By adding additional
points on two hemispherical caps at the ends of the cylinder, we can
extend $S$ into a uniform sample of a smooth convex surface with
bounded curvature and constant local feature size.

Let $h_\alpha(t)$ denote the helix $(\alpha t, \cos t, \sin t)$, where
${\alpha>0}$ is a fixed parameter called the \emph{pitch}.  Using
elementary trigonometric identities and matrix operations, we can
simplify the insphere determinant for five points on this helix as
follows.
\begin{multline*}
	\begin{vmatrix}
	1 & \alpha t_1 & \cos t_1 & \sin t_1 & \alpha^2 t_1^2 + \cos^2 t_1 + \sin^2 t_1 \\
	1 & \alpha t_2 & \cos t_2 & \sin t_2 & \alpha^2 t_2^2 + \cos^2 t_2 + \sin^2 t_2 \\
	1 & \alpha t_3 & \cos t_3 & \sin t_3 & \alpha^2 t_3^2 + \cos^2 t_3 + \sin^2 t_3 \\
	1 & \alpha t_4 & \cos t_4 & \sin t_4 & \alpha^2 t_4^2 + \cos^2 t_4 + \sin^2 t_4 \\
	1 & \alpha t_5 & \cos t_5 & \sin t_5 & \alpha^2 t_5^2 + \cos^2 t_5 + \sin^2 t_5 \\
	\end{vmatrix}
\\	=
	\alpha^3
	\begin{vmatrix}
	1 & t_1 & \cos t_1 & \sin t_1 & t_1^2 \\
	1 & t_2 & \cos t_2 & \sin t_2 & t_2^2 \\
	1 & t_3 & \cos t_3 & \sin t_3 & t_3^2 \\
	1 & t_4 & \cos t_4 & \sin t_4 & t_4^2 \\
	1 & t_5 & \cos t_5 & \sin t_5 & t_5^2 \\
	\end{vmatrix}
\end{multline*}
We obtain the surprising observation that changing the pitch $\alpha$
of the helix does not change the combinatorial structure of the
Delaunay triangulation of any set of points on the helix.  (More
generally, scaling any set of points on any circular cylinder along
the cylinder's axis leaves the Delaunay triangulation invariant.)
Thus, for purposes of analysis, it suffices to consider the case
$\alpha=1$.  Let $h(t) = h_1(t) = (t, \cos t, \sin t)$.

Our first important observation is that any set of points on a single
turn of any helix has a \emph{neighborly} Delaunay triangulation,
meaning that every pair of points is connected by a Delaunay edge. For
any real value $t$, we define the \emph{bitangent sphere} $\beta(t)$
as the unique sphere passing through $h(t)$ and $h(-t)$ and tangent to
the helix at those two points.

\begin{lemma}
\label{L:bitangent}
For any $0<t<\pi$, the sphere $\beta(t)$ intersects the helix $h$ only
at its two points of tangency.
\end{lemma}

\begin{proof}
Symmetry considerations imply that the bitangent sphere must be
centered on the $y$-axis, so it can be described by the equation $x^2
+ (y-a)^2 + z^2 = r^2$ for some constants $a$ and~$r$.  Let $\gamma$
denote the intersection curve of $\beta(t)$ and the cylinder $y^2 +
z^2 = 1$.  Every intersection point between $\beta(t)$ and the helix
must lie on $\gamma$.  If we project the helix and the intersection
curve to the $xy$-plane, we obtain the sinusoid $y = \cos x$ and a
portion of the parabola $y = \gamma(x) = (x^2 - r^2 + a^2 + 1)/2a$.
These two curves meet tangentially at the points $(t, \cos t)$ and
$(-t, \cos t)$.

\begin{figure}[htb]
    \centerline{\epsfig{file=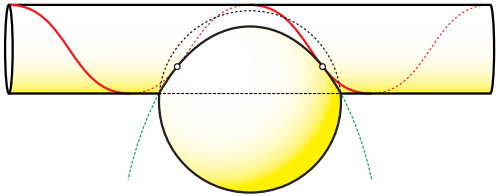,height=1in}}
    \caption{The intersection curve of the cylinder and a bitangent 
    sphere projects to a parabola on the $xy$-plane.}
\end{figure}

The mean value theorem implies that $\gamma(x) = \cos x$ at most four
times in the range $-\pi < x < \pi$.  (Otherwise, the curves $y'' =
-\cos x$ and $y'' = \gamma''(x) = 1/a$ would intersect more than twice
in that range.)  Since the curves meet with even multiplicity at two
points, those are the only intersection points in the range $-\pi < x
< \pi$.  Since $\gamma(x)$ is concave, we have $\gamma(\pm\pi) <
\cos\pm\pi = -1$, so there are no intersections with $\abs{x}\ge\pi$.
Thus, the curves meet only at their two points of tangency.
\end{proof}

\begin{corollary}
\label{C:spring}
Any set $S$ of $n$ points on the helix $h(t)$ in the range $-\pi < t <
\pi$ has a neighborly Delaunay triangulation.
\end{corollary}

\begin{proof}
Let $p$ and $q$ be arbitrary points in $S$, and let~$\beta$ be the
unique ball tangent to the helix at $p$ and $q$.  By
Lemma~\ref{L:bitangent}, $\beta$ does not otherwise intersect the
helix and therefore contains no point in~$S$.  Thus, $p$ and $q$ are
neighbors in the Delaunay triangulation of $S$.
\end{proof}

We can now easily complete the analysis of our helical point set
$S_{\!\sqrt{n}}$.  Lemma \ref{L:bitangent} implies that every point in
$S_{\!\sqrt{n}}$ is connected by a Delaunay edge to every other point
less than a full turn around the helix
$h_{\sqrt{\vphantom{t}\smash{1/n}}}(t)$, and each full turn of the
helix contains $\floor{2\pi\sqrt{n}}$ points.  Thus, the number of
edges in the Delaunay triangulation of $S_{\!\sqrt{n}}$ is at least
$2\pi n^{3/2} - \Theta(n)$.

\begin{theorem}
For any $n$, there is a set of $n$ points in $\Real^3$ with spread 
$\sqrt{n}$ whose Delaunay triangulation has complexity 
$\Omega(n^{3/2})$.  Moreover, this point set is a uniform sample of a 
smooth convex surface with constant local feature size.
\end{theorem}

We can generalize our helix construction to other values of the spread
$\Delta$ as follows.

\begin{theorem}
\label{Th:lower}
For any $n$ and $\Delta = \Omega(n^{1/3})$, there is a set of $n$ 
points in $\Real^3$ with spread $\Delta$ whose Delaunay triangulation 
has complexity $\Omega(\min\set{\Delta^3, n\Delta, n^2})$.
\end{theorem}

\begin{proof}
There are three cases to consider, depending on whether the spread is
at least $n$, between $\sqrt{n}$ and $n$, or at most $\sqrt{n}$.  The
first case is trivial.  For the case ${\sqrt{n} \le \Delta \le n}$, we
take a set of evenly spaced points on a helix with pitch $\Delta/n$:
\[
	S_\Delta =
	\Setbar{\Paren{\frac{t}{n},\,
			\cos\frac{t}{\Delta},\,
			\sin\frac{t}{\Delta}}}
		{t \in [n]}.
\]
Every point in $S_\Delta$ is connected by a Delaunay edge to every
other point less than a full turn away on the helix, and each turn of
the helix contains $\Omega(\Delta)$ points, so the total complexity of
the Delaunay triangulation is $\Omega(n\Delta)$.

The final case $n^{1/3} \le \Delta \le \sqrt{n}$ is slightly more
complicated.  Our point set consists of several copies of
our helix construction, with the helices positioned at the points of a
square lattice, so the entire construction loosely resembles a
mattress.  Specifically, $S_\Delta$ is the set
\[
	\Setbar{\Paren{\frac{t}{r},\,
	 		4i+\cos\frac{t}{\sqrt{r}},\,
			4j+\sin\frac{t}{\sqrt{r}}}}
		{t \in [wr];\, i,j \in [w]},
\]
where $r$ and $w$ are parameters to be determined shortly.  This set
contains $n = w^3 r$ points.  The diameter of $S_\Delta$ is
$\Theta(w)$ and the closest pair distance is $\Theta(1/\sqrt{r})$, so
its spread is $\Delta = \Theta(w\sqrt{r})$.  Thus, given $n$ and
$\Delta$, we have $w = \Theta(n/\Delta^2)$ and $r =
\Theta(\Delta^6/n^2)$.  Straightforward calculations imply that for
all $t<\pi/4$ and $\alpha<1$, the bitangent sphere $\beta_\alpha(t)$
has radius less than $2$.  Since adjacent helices are separated by
distance~$2$, every point in $S_\Delta$ is connected in the Delunay
triangulation to every point at most half a turn away in the same
helix.  Each turn of each helix contains $\Omega(\sqrt{r})$ points, so
the Delaunay triangulation of $S_\Delta$ has complexity
$\Omega(n\sqrt{r}) = \Omega(\Delta^3)$.
\end{proof}

\subsection{Upper Bounds}
\label{SS:upper}

\begin{figure}[htb]
\centerline{\epsfig{file=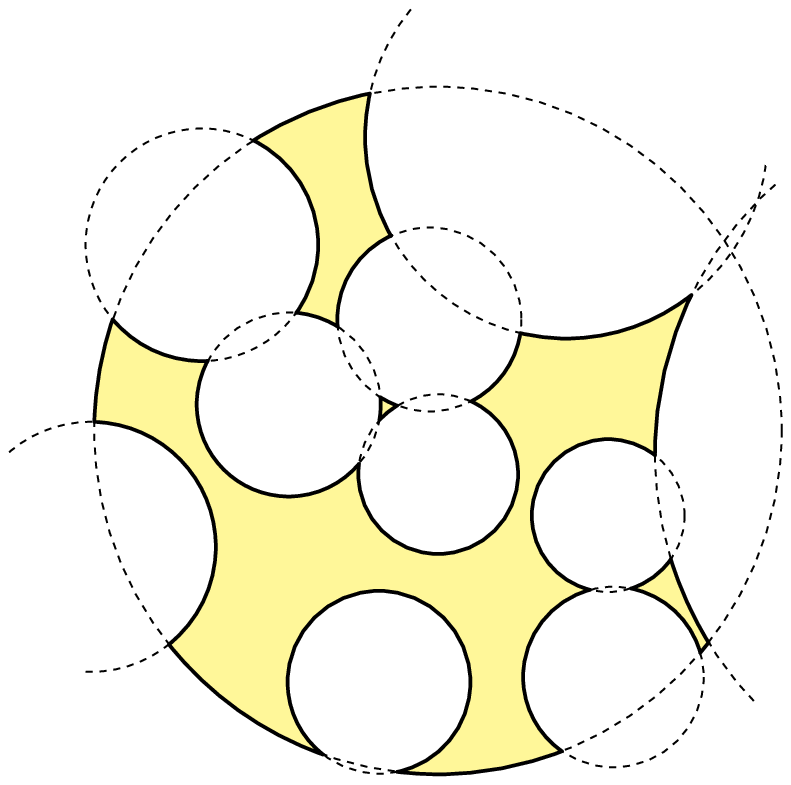,height=2in}}
\caption{Swiss cheese (in $\Real^2$)}
\label{Fig:cheese}
\end{figure}

Let $B$ be a ball of radius $R$ in $\Real^3$, and let $b_1, b_2, b_3,
\dots$ be balls of radius at least $r$, where $1\le r\le R$.  Our
upper bound proof uses the following geometric properties of the
`Swiss cheese' $C = B \setminus \bigcup_i b_i$.  See
Figure~\ref{Fig:cheese}.

\begin{lemma}
\label{L:cheese}
The surface area of $C$ is $O(R^3/r)$.
\end{lemma}

\begin{proof}
The outer surface $\partial C\cap\partial B$ clearly has area $O(R^2)
= O(R^3/r)$, so it suffices to bound the surface area of the `holes'.
For each $i$, let $H_i = B\cap \partial b_i$ be the boundary of the
$i$th hole, and let $H = \bigcup_i H_i = \partial C \setminus \partial
B$.  For any point $x\in H$, let $s_x$ denote the open line segment of
length $r$ extending from $x$ towards the center of the ball $b_i$
with $x$ on its boundary.  (If $x$ lies on the surface of more than
one $b_i$, choose one arbitrarily.)  Let $S = \bigcup_{x\in H} s_x$ be
the union of all such segments, and for each $i$, let $S_i =
\bigcup_{x\in H_i} s_x$.  Each $S_i$ is a fragment of a spherical
shell of thickness $r$ inside the ball $b_i$.  See
Figure~\ref{Fig:shell}.

\begin{figure}[htb]
\centerline{\epsfig{file=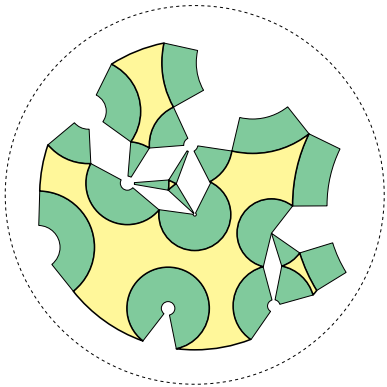,height=2in}}
\caption{Shell fragments used to bound the surface area of $C$.}
\label{Fig:shell}
\end{figure}

For each $i$, we have (after some tedious calculations)
\[
	\vol(S_i) = \left( \frac{r_i}{3}
		    - \frac{(r-r_i)^3}{3 r_i^2} \right) \area(H_i) 
		  \ge \frac{r}{3}\area(H_i),
\]
where $r_i \ge r$ is the radius of $b_i$.  The triangle inequality
implies that $s_x$ and $s_y$ are disjoint for any two points $x,y\in
H$, so the shell fragments $S_i$ are pairwise disjoint.  Finally,
since $S$ fits inside a ball of radius $R+r \le 2R$, its volume is
$O(R^3)$.  Thus, $\area(H) = \sum_i \area(H_i) \le \sum_i 3\vol(S_i)/r
= 3\vol(S)/r = O(R^3/r)$.
\end{proof}

\begin{lemma}
\label{L:eat}
Let $U$ be any unit ball whose center is in $C$ and at distance $2/3$
from $\partial C$.  Then $U$ contains $\Omega(1)$ surface area of~$C$.
\end{lemma}

\begin{figure}[htb]
\centerline{\epsfig{file=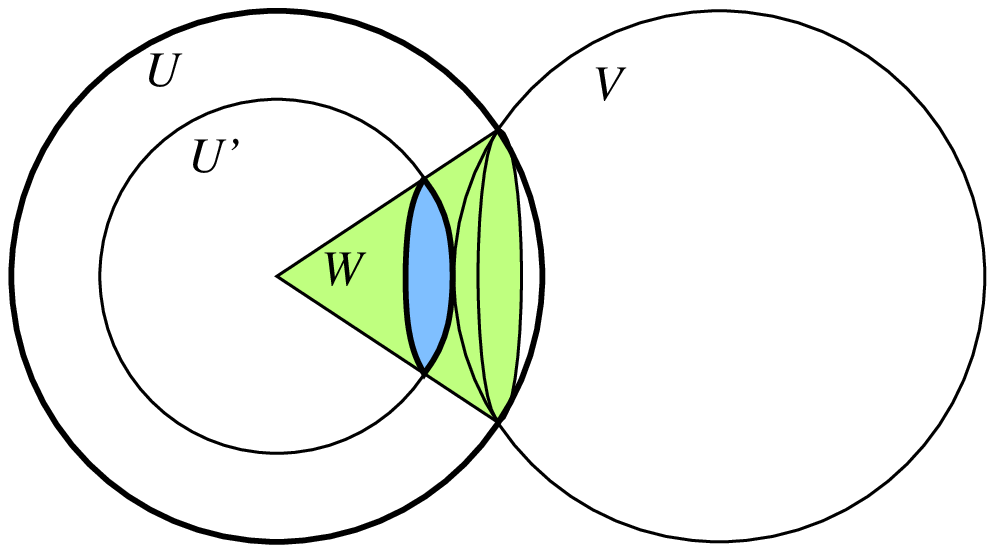,height=1.2in}}
\caption{Proof of Lemma \ref{L:eat}}
\label{Fig/eat}
\end{figure}

\begin{proof}
Without loss of generality, assume that $U$ is centered at the origin
and that $(0, 0, 2/3)$ is the closest point of $\partial C$ to the
origin.  Let $U'$ be the open ball of radius $\delta$ centered at the
origin, let $V$ be the open unit ball centered at $(0, 0, 5/3)$, and
let $W$ be the cone whose apex is the origin and whose base is the
circle $\partial U\cap \partial V$.  See Figure \ref{Fig/eat}.  $U'$
lies entirely inside $C$, and since $r\ge 1$, we easily observe that
$V$ lies entirely outside $C$.  Thus, the surface area of $\partial C
\cap W \subseteq \partial C \cap U$ is at least the area of the
spherical cap $\partial U' \cap W$, which is exactly $4\pi/27$.
\end{proof}
\unskip

\begin{theorem}
\label{Th:degree}
Let $S$ be a set of points in $\Real^3$ whose closest pair is at 
distance $2$, and let $r$ be any real number.  Any point in $S$
has $O(r^2)$ Delaunay neighbors at distance at most $r$.
\end{theorem}

\begin{proof}
Let $o$ be an arbitrary point in $S$, and let $B$ be a ball of radius 
$r$ centered at $o$.  Call a Delaunay neighbor of $o$ a \emph{friend} 
if it lies inside $B$, and call a friend $q$ \emph{interesting} if 
there is another point $p\in S$ (not necessarily a Delaunay neighbor 
of $o$) such that $\abs{op} < \abs{oq}$ and $\angle poq < 1/r$.  A 
simple packing argument shows that $o$ has at most $O(r^2)$ boring 
friends.

Let $Q$ be the set of interesting friends of~$o$.  Every point $q\in
Q$ lies on the boundary of a Delaunay ball $d_q$ that contains no
points of $S$ in its interior and also has $o$ on its boundary.  It is
straightforward to prove that because $q$ is interesting and has
distance at least $2$ from any other point, $d_q$ must have radius at
least $r$.  Let $b_q$ be the ball concentric with $d_q$ with radius
$2/3$ less than the radius of $d_q$.  Finally, for any point $q$, let
$U_q$ be the unit-radius ball centered at $q$.

We now have a set of unit balls, one for each interesting friend of
$o$, whose centers lie at distance exactly $2/3$ from the boundary of
the Swiss cheese $C = B \setminus \bigcup_{q\in Q} b_q$.  By Lemma
\ref{L:cheese}, $C$ has surface area $O(r^2)$, and by Lemma
\ref{L:eat}, each unit ball $U_q$ contains $\Omega(1)$ surface area
of~$C$.  Since the unit balls are disjoint, it follows that $o$ has at
most $O(r^2)$ interesting friends.
\end{proof}

\begin{theorem}
\label{Th:reach}
Let $S$ be a set of points in $\Real^3$ whose closest pair is at 
distance $2$ and whose diameter is $2\Delta$, and let $r$ be any real 
number.  There are $O(\Delta^3/r)$ points in $S$ with a Delaunay 
neighbor at distance at least $r$.
\end{theorem}

\begin{proof}
Call a point \emph{far-reaching} if it has a Delaunay neighbor at 
distance at least $r$, and let $Q$ be the set of far-reaching points.  
Let $B$ be a ball of radius $2\Delta$ containing $S$.  For each $q\in 
Q$, let $f_q$ be a maximal empty ball containing $q$ and its furthest 
Delaunay neighbor, and let $b_q$ be the concentric ball with radius 
$2/3$ smaller than $f_p$.  By construction, each ball $b_q$ has radius 
at least $r/2 - 2/3$.  Finally, for any far-reaching point $q$, let 
$U_q$ be the unit-radius ball centered at $q$.  By Lemma 
\ref{L:cheese}, the Swiss cheese $C = B \setminus \bigcup_{q\in Q} 
b_q$ has surface area $O(\Delta^3/r)$, and by Lemma \ref{L:eat}, each 
unit ball $U_q$ contains $\Omega(1)$ surface area of~$C$.  Since these 
unit balls are disjoint, there are at most $O(\Delta^3/r)$ of them.
\end{proof}

\begin{corollary}
\label{C:upper}
Let $S$ be a set of points in $\Real^3$ with spread~$\Delta$.  The
Delaunay triangulation of $S$ has complexity $O(\Delta^4)$.
\end{corollary}

\begin{proof}
For all $r$, let $F(r)$ be the number of far-reaching points in $S$,
\ie, those with Delaunay edges of length at least $r$.  From
Theorem~\ref{Th:reach}, we have $F(r) = O(\Delta^3/r)$.  By
Theorem~\ref{Th:degree}, if the farthest neighbor of a point $p$ is at
distance between $r$ and $r+1$, then $p$ has $O(r^2)$ neighbors.
Thus, the total number of Delaunay edges is at most
\begin{align*}
	\sum_{r=0}^{\Delta} O(r^2)\cdot \big(F(r) - F(r+1)\big)
	&=
	\sum_{r=0}^{\Delta} O(r)\cdot F(r)
\\	&=
	\sum_{r=0}^{\Delta} O(\Delta^3)
\\	&=
	O(\Delta^4)
\end{align*}
\aftermath
\end{proof}

\subsection{Conjectured Upper Bounds}
\label{SS:conj}

I conjecture that the lower bounds in Theorem \ref{Th:lower} are
tight, but Corollary \ref{C:upper} is the best upper bound known.
Nearly matching upper bounds could be derived from the following
conjecture using a divide and conquer argument, suggested by Edgar
Ramos (personal communication).

Let $S$ be a \emph{well-separated} set of points with closest pair
distance~$1$, lying in two balls of radius $\Delta$ that are separated
by distance at least $c\Delta$ for some constant ${c>1}$.  Call an
edge in the Delaunay triangulation of~$S$ a \emph{crossing edge} if it
has one endpoint in each ball.

\begin{conjecture}\label{split?}
Some point in $S$ is an endpoint of $O(\Delta)$ crossing edges.
\end{conjecture}

\begin{lemma}\label{L:split}
Conjecture \ref{split?} implies that the Delaunay triangulation of $S$
has $O(\min\set{\Delta^3, \Delta n, n^2})$ crossing edges.
\end{lemma}

\begin{proof}
Theorem~\ref{Th:reach} implies that only $O(\Delta^2)$ points can be
endpoints of crossing edges.  Thus, we can assume without loss of
generality that $n = O(\Delta^2)$.

We compute the total number of crossing edges by iteratively removing
the point with the fewest crossing edges and retriangulating the
resulting hole, say by incremental flipping.  Conjecture \ref{split?}
implies that we delete only $O(\Delta)$ crossing edges with each
point, so altogether we delete $O(n\Delta) = O(\Delta^3)$ crossing
edges.  Not all of these edges are in the original Delaunay
triangulation, but that only helps us.
\end{proof}

\begin{theorem}
Conjecture \ref{split?} implies that the Delaunay triangulation of $n$ 
points in $\Real^3$ with spread $\Delta$ has complexity $O(\min 
\set{\Delta^3 \log \Delta, n \Delta, n^2})$.
\end{theorem}

\begin{proof}
Assume Conjecture \ref{split?} is true, and let $S$ be an arbitrary
set of $n$ points with diameter~$\Delta$, where the closest pair of
points is at unit distance.  $S$ is contained in an axis-parallel cube
$C$ of width $\Delta$.  We construct a \emph{well-separated pair
decomposition} of $S$ \cite{ck-dmpsa-95}, based on a simple octtree
decomposition of $C$.  The octtree has $O(\log \Delta)$ levels.  At
each level $i$, there are $8^i$ cells, each a cube of width
$\Delta/2^i$.  Our well-separated pair decomposition includes, for
each level~$i$, the points in any pair of level-$i$ cells separated by
a distance between $c\Delta/2^i$ and $2c\Delta/2^i$.  A simple packing
argument implies that any cell in the octtree is paired with $O(1)$
other cells, all at the same level, and so any point appears in
$O(\log \Delta)$ subset pairs.  Every Delaunay edge of $S$ is a
crossing edge for some well-separated pair of cells.

Lemma \ref{L:split} implies that the points in any well-separated pair
of level-$i$ cells have $O(\Delta^3/8^i)$ crossing Delaunay edges.
Since there are $O(8^i)$ such pairs, the total number of crossing
edges between level-$i$ cells is $O(\Delta^3)$.  Thus, there are
$O(\Delta^3\log \Delta)$ Delaunay edges altogether.

Lemma \ref{L:split} also implies that for any well-separated pair of
level-$i$ cells, the average number of crossing edges per point is
$O(\Delta/2^i)$.  Since every point belongs to a constant number of
subset pairs at each level, the total number of crossing edges at
level $i$ is $O(n\Delta/2^i)$.  Thus, the total number of Delaunay
edges is $O(n\Delta)$.
\end{proof}

This upper bound is still a logarithmic factor away from our lower
bound construction when $\Delta = o(\sqrt{n})$.  However, our argument
is quite conservative; all crossing edges for a well-separated pair of
subsets are counted, even though some or all of these edges may be
blocked by other points in $S$.  A more careful analysis would
probably eliminate the final logarithmic factor.

\section{Nice Surface Data}
\label{S:surface}

\noindent
Let $\Sigma$ be a smooth surface without boundary in $\Real^3$.  The
\emph{medial axis} of $\Sigma$ is the closure of the set of points in
$\Real^3$ that have more than one nearest neighbor on $\Sigma$.  The
\emph{local feature size} of a point $x\in\Sigma$, denoted $\lfs(x)$,
is the distance from $x$ to the medial axis of $\Sigma$.  Let~$S$ be a
set of \emph{sample points} on~$\Sigma$.  Following Amenta and
Bern~\cite{ab-srvf-99}, we say that $S$ is an \emph{$\e$-sample}
of~$\Sigma$ if the distance from any point $x\in\Sigma$ to the
nearest sample point is at most $\e\cdot\lfs(x)$.

The first step in several surface reconstruction algorithms is to
construct the Delaunay triangulation or Voronoi diagram of the sample
points.  Edelsbrunner and \Mucke~\cite{em-tdas-94} and Bajaj \etal\
\cite{bbx-arssf-95,bb-srmua-97} describe algorithms based on
\emph{alpha shapes}, which are subcomplexes of the Delaunay
triangulation; see also \cite{gmw-sruas-97}.  Extending earlier work
on planar curve reconstruction \cite{abe-cbscc-98,g-cac-99}, Amenta
and Bern \cite{ab-srvf-99,abk-nvbsr-98} developed an algorithm to
extract a certain manifold subcomplex of the Delaunay triangulation,
called the \emph{crust}.  Amenta \etal~\cite{acdl-sahsr-00} simplified
the crust algorithm and proved that if $S$ is an $\e$-sample of a
smooth surface $\Sigma$, for some sufficiently small~$\e$, then the
crust is homeomorphic to $\Sigma$.  Boissonnat and
Cazals~\cite{bc-ssrnn-00} and Hiyoshi and Sugihara \cite{hs-vbihc-00}
proposed algorithms to produce a smooth surface using natural
coordinates, which are defined and computed using the Voronoi diagram
of the sample points.  Further examples can be found in
\cite{ack-pcubm-,a-rrsru-98,b-r2d3d-84,cdes-dst-01}.

In this section, we show that $\e$-samples of smooth surfaces can have
complicated Delaunay triangulations, implying that all these surface
reconstruction algorithms can take quadratic time in the worst case.
We will analyze our constructions in terms of the \emph{sample
measure} of a surface $\Sigma$, which we define as follows:
\[
	\mu(\Sigma) = \int_\Sigma \frac{dx}{\lfs^2(x)}.
\]

\begin{lemma}
For all $\e<1/2$, every $\e$-sample of $\Sigma$ contains
$\Omega(\mu(\Sigma)/\e^2)$ points.
\end{lemma}

\begin{proof}
Let $S$ be an arbitrary $\e$-sample of $\Sigma$.  Amenta and
Bern~\cite{ab-srvf-99} observed that $\abs{\lfs(p) - \lfs(q)} <
\abs{pq}$ for any points $p, q \in \Sigma$.  This observation implies
that for any point $x\in \Sigma$, we have $\abs{xp} \le \e \lfs(x) \le
\e({\lfs(p) + \abs{xp}}) \le \frac{\e}{1-\e} \lfs(p)$, where $p\in S$
is the sample point closest to~$x$.  Thus, we can cover $\Sigma$ with
circular neighborhoods of radius $\frac{\e}{1-\e}\lfs(p)$ around each
sample point ${p\in S}$.  By similar arguments, the neighborhood of
$p$ has area at least ${\pi(\frac{\e}{1-\e} (1 - \frac{\e}{1-\e})
\lfs(p))^2}$, and any point in the neighborhood of $p$ has local
feature size at most ${(1+\frac{\e}{1-\e})}\lfs(p)$.  It follows that
each neighborhood has sample measure $\Omega(\e^2)$, and since there
are $n$ such neighborhoods, $\mu(\Sigma) = O(n\e^2)$.
\end{proof}

We say that an $\e$-sample is \emph{parsimonious} if it contains
$O(\mu(\Sigma)/\e^2)$ points.

\subsection{Oversampling Is Bad}
\label{SS:over}

The easiest method to produce a surface sample with high Delaunay
complexity is \emph{oversampling}, where some region of the surface
contains many more points than necessary.  In fact, the only surface
where oversampling cannot produce a quadratic-complexity Delaunay
triangulation is the sphere, even if we only consider parsimonious
samples.

\begin{theorem}
For any smooth non-spherical surface~$\Sigma$, any $\e>0$, and any
sufficiently large $n$, there is a parsimonious $\e$-sample of
$\Sigma$ of size $n$ whose Delaunay triangulation has complexity
$\Omega(n^2)$.
\end{theorem}

\begin{proof}
Let $S$ be any parsimonious $\e$-sample of $\Sigma$.  Let~$\sigma$ be
a small sphere intersecting~$\Sigma$ in a non-planar curve, where the
distance from $\sigma$ to any point os $S$ is at elast the radius of
$\sigma$.  Such a sphere always exists unless $\Sigma$ is itself a
sphere.  Let $\alpha$ and~$\beta$ be extremely short segments of the
intersection curve ${\Sigma\cap\sigma}$ that approximate skew line
segments.  Straighten these curves slightly, keeping them on the
surface $\Sigma$ and keeping the endpoints fixed, to obtain curves
$\alpha'$~and~$\beta'$.  Finally, let $A$ and $B$ be sets of $\abs{S}$
evenly spaced points on $\alpha'$ and $\beta'$, respectively.  See
Figure~\ref{Fig:oversample}.

\begin{figure}[htb]
\centerline{\epsfig{file=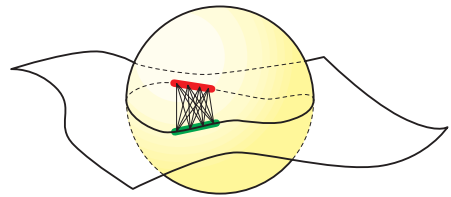,height=1in}}
\caption{Parsimoniously oversampling a non-spherical surface.}
\label{Fig:oversample}
\end{figure}

The Delaunay triangulation of $A\cup B$ has complexity
$\Omega(\abs{S}^2)$; every point in $A$ is a Delaunay neighbor of
every point in $B$.  Moreover, any Delaunay circumsphere of $A\cup B$
closely approximates the sphere $\sigma$ and thus excludes every point
in $S$.  Thus, $S\cup A\cup B$ is a parsimonious $\e$-sample of
$\Sigma$ consisting of $n = 3\abs{S}$ points whose Delaunay
triangulation has complexity $\Omega(n^2)$.
\end{proof}

The reconstruction algorithm of Amenta~\etal~\cite{acdl-sahsr-00}
extracts a surface from a subset of the Delaunay triangles of the
sample points.  Their algorithm estimates the surface normal at each
smple point $p$ using the Voronoi diagram of the samples.  The
\emph{cocone} at $p$ is the complement of a very wide double cone
whose apex is $p$ and whose axis is the estimated normal vector
at~$p$.  The algorithm extracts the Delaunay triangles whose dual
Delaunay edges intersect the cocones of all three of its vertices, and
then extracts a manifold surface from those cocone triangles.  Usually
only a small subset of the Delaunay triangles pass this filtering
phase, but our construction shows that there can be $\Omega(n^2)$
cocone triangles in the worst case.

\subsection{Uniform Sampling Can Still Be Bad}
\label{SS:wedge}

Unfortunately, oversampling is not the only way to get quadratic
Delaunay triangulations.  Let $S$ be a set of sample points on the
surface $\Sigma$.  We define the \emph{second sampling density} of a
point $x \in \Sigma$, denoted $\sd_2(x)$, as the distance from $x$ to
the second closest sample point, divided by $\lfs(x)$.  We say that
$S$ is a \emph{uniform $\e$-sample} of $\Sigma$ if $\e/4 \le \sd_2(x)
\le \e$ for all $x\in\Sigma$.%
\footnote{There is nothing special about the number $4$ here; any
constant $c>2$ will do.  However, as $c$ approaches $2$, the maximum
$\e$ for which a $c$-uniform $\e$-sample exists approaches zero.}
Uniform $\e$-samples are also parsimonious $\e$-samples, but with
absolutely \emph{no} oversampling.  In particular, the size of any
uniform $\e$-sample is $\Theta(\mu(\Sigma)/\e^2)$.

\begin{lemma}
\label{L:sausages}
For any $n$ and $\e > \sqrt{1/n}$, there is a two-component
surface~$\Sigma$ and an $n$-point uniform $\e$-sample~$S$ of $\Sigma$,
such that the Delaunay triangulation of $S$ has complexity
$\Omega(n^2\e^2)$.
\end{lemma}

\begin{proof}
The surface $\Sigma$ is the boundary of two \emph{sausages} $\Sigma_x$
and $\Sigma_y$, each of which is the Minkowski sum of a unit sphere
and a line segment.  Specifically, let
\begin{align*}
	\Sigma_x &= U + \overline{(-w,0,d+1), (w,0,d+1)}\qquad\text{and}
\\
	\Sigma_y &= U + \overline{(0,-w,-d-1), (0,w,-d-1)},
\end{align*}
where $U$ is the unit ball centered at the origin, ${w = n\e^2}$, and
$d = 4w/\e = 4n\e$.  The local feature size of every point on $\Sigma$
is $1$, so any uniform $\e$-sample of $\Sigma$ has ${\Theta((w+1)/\e^2)}
= \Theta(n)$ points.

Define the \emph{seams} $\sigma_x$ and $\sigma_y$ as the maximal line
segments in each sausage closest to the $xy$-plane:
\begin{align*}
	\sigma_x &= \overline{(-w,0,d), (w,0,d)} \qquad\text{and}
\\	\sigma_y &= \overline{(0,-w,-d), (0,w,-d)}.
\end{align*}
Our uniform $\e$-sample $S$ contains $2w/\e+1$ points along each seam:
\begin{alignat*}{2}
	p_i &= (i\e, 0, d)
	    &\quad \text{for all integers~}
	    & {-w}/\e \le i\le w/\e,
\text{~and}
\\	q_j &= (0, j\e, -d)
	    &\quad \text{for all integers~}
	    & {-w}/\e \le j\le w/\e.
\end{alignat*}
The Delaunay triangulation of these $\Theta(w/\e) = \Theta(n\e)$
points has complexity $\Theta(w^2/\e^2) = \Theta(n^2\e^2)$.

Let $\gamma_{ij}$ be the ball whose boundary passes through $p_i$ and
$q_j$ and is tangent to both seams.  This ball may contain other
portions of the surface, but we claim that the intersection is small
enough that we can avoid it with our sample points.  The intersection
of $\Sigma_x$ and~$\gamma_{ij}$ is a small oval, tangent to~$p_i$ and
symmetric about the plane $x=i\e$.  Tedious calculation (which we
omit) implies that the width of the oval is
\[
	2\tan^{-1} \left(\frac{4dj\e}{4d(d+1) + (i^2-j^2)\e^2}\right)
	< \frac{4w}{d}
	= \e.
\]
See Figure \ref{Fig:sausages}.

\begin{figure}[htb]
\centerline{\epsfig{file=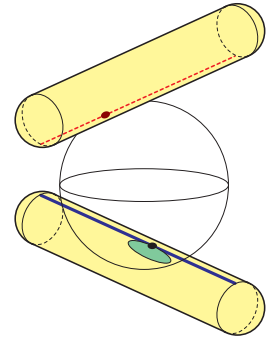,height=1.75in}}
\caption{Two sausages and a sphere tangent to both seams.}
\label{Fig:sausages}
\end{figure}

So $\Sigma_x\cap\gamma_{ij}$ lies entirely within a strip of width
$2\e$ centered along the seam $\sigma_x$.  A symmetric argument gives
the analogous result for $\Sigma_y\cap\gamma_{ij}$.  We can uniformly
sample $\Sigma$ so that no other sample point lies within either
strip.  Each segment $\overline{p_iq_j}$ is an edge in the Delaunay
triangulation of the sample, and there are $\Omega(w^2/\e^2) =
\Omega(n^2\e^2)$ such segments.
\end{proof}

\begin{theorem}
\label{Th:wedgie}
For any $n$ and any $\e > \sqrt{(\log n)/n}$, there is a connected
surface $\Sigma$ and an $n$-point uniform $\e$-sample $S$ of $\Sigma$,
such that the Delaunay triangulation of $S$ has complexity
$\Omega(n^2\e^2)$.
\end{theorem}

\begin{proof}
Intuitively, we produce the surface $\Sigma$ by pushing two sausages
into a spherical balloon.  These sausages create a pair of conical
\emph{wedges} inside the balloon whose \emph{seams} lie along two skew
lines.  The local feature size is small near the seams and drops off
quickly elsewhere, so a large fraction of the points in any uniform
sample must lie near the seams.  We construct a particular sample with
points \emph{exactly} along the seams that form a quadratic-complexity
triangulation, similarly to our earlier sausage construction.  Our
construction relies on several parameters: the radius~$R$ of the
spherical balloon, the width~$w$ and height~$h$ of the wedges, and the
distance $d$ between the seams.

\begin{figure}[htb]
\centerline{\epsfig{file=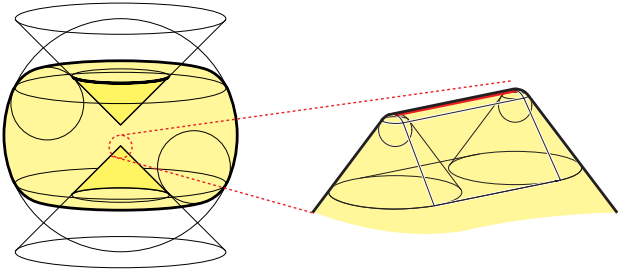,height=1.4in}}
\caption{A smooth surface with a bad uniform $\e$-sample, and a
closeup of one of its wedges.}
\label{Fig:bagel}
\end{figure}

Each wedge is the Minkowski sum of a unit sphere, a right circular
cone with height $h$ centered along the $z$-axis, and a line segment
of length $w$ parallel to one of the other coordinate axes.  The
boundary of each wedge can be decomposed into cylindrical, spherical,
conical, and planar facets.  The cylindrical and spherical facets
constitute the \emph{blade} of the wedge, and the \emph{seam} of the
blade is the line segment of length $w$ that bisects the cylindrical
facet.  The local feature size of any point on the blade is exactly
$1$, and the local feature size of any other boundary point is its
distance from the blade.  Straightforward calculations imply that the
sample measure of the wedge is $O(w + \log h + 1)$.

A first approximation $\widetilde{\Sigma}$ of the surface $\Sigma$ is
obtained by removing two wedges from a ball of radius $R$ centered at
the origin.  One wedge points into the ball from below; its seam is
parallel to the $x$-axis and is centered at the point $(0,0,-R+h)$.
The other wedge points into the ball from above; its seam is parallel
to the $y$-axis and is centered at $(0,0,R-h)$.  Let $d = 2R-2h-2$
denote the distance between the wedges.  Our construction has $1\ll
w\ll d\ll h$, so $R < 3h$.

To obtain the final smooth surface $\Sigma$, we round off the sharp
edges by rolling a ball of radius $h/4$ inside~$\widetilde{\Sigma}$
along the wedge/balloon intersection curves.  We call the resulting
warped toroidal patches the \emph{sleeves}.  The local feature size of
any point on the sleeves or on the balloon is at least $h/4$.  Since
$\Sigma$ is star-shaped and contained in a sphere of radius $R$, its
surface area is at most $4\pi R^2 < 36\pi h^2$.  It follows that the
sleeves have constant sample measure.  The local feature size of wedge
points changes only far from the blades and by only a small constant
factor, so $\mu(\Sigma) = \Theta(w + \log h + 1)$.  To complete the
construction, we set $w = n\e^2$, $d = 4n\e$, and $h = 20n\e$.  See
Figure~\ref{Fig:bagel}.

Finally, we construct a uniform $\e$-sample $S$ with $\Theta(w/\e)$
sample points evenly spaced along each seam and every other point at
least $\e$ away from the seams.  Setting $h > 5d$ (and thus $R > 10d$)
ensures that the Delaunay spheres $\gamma_{ij}$ between seam points do
not touch the surface except on the blades.  By the argument in Lemma
\ref{L:sausages}, there are $\Omega(w^2/\e^2) = \Omega(n^2\e^2)$
Delaunay edges between seam points.
\end{proof}

\subsection{Some Surfaces Are Just Evil}
\label{SS:teeth}

In this section, we describe a family of surfaces for which \emph{any}
parsimonious $\e$-sample has a Delaunay triangulation of
near-quadratic complexity.  First we give a nearly trivial
construction of a bad surface with several components, and then we
join these components into a single connected surface using a method
similar to Theorem \ref{Th:wedgie}.

\begin{lemma}
\label{L:balls}
For any $n$ and any $\e < \sqrt{1/n}$, there is a smooth
surface~$\Sigma$ such that the Delaunay triangulation of any
parsimonious $\e$-sample of $\Sigma$ has complexity $\Omega(n^2\e^4)$,
where $n$ is the size of the sample.
\end{lemma}

\begin{proof}
Let $P$ be a set containing the following $k$ points:
\begin{alignat*}{2}
	p_i &= (ik, 0, k^2)
	    &\quad \text{for all integers~}
	    & {-k/4} \le i\le k/4,
\text{~and}
\\	q_j &= (0, jk, -k^2)
	    &\quad \text{for all integers~}
	    & {-k/4} \le j\le k/4.
\end{alignat*}
We easily verify that every pair of points $p_i$ and $q_j$ lie on a
sphere $\gamma_{ij}$ with every other point in $P$ at least unit
distance outside.

Let $\Sigma = \bigcup_{p\in P} U_p$, where $U_p$ is the unit-radius
sphere centered at $p$.  Clearly, $\lfs(x)=1$ for every point
$x\in\Sigma$, so $\mu(\Sigma) = 4\pi k$.  Let $S$ be an arbitrary
parsimonious $\e$-sample of $\Sigma$, let $n = \abs{S} =
\Theta(k/\e^2)$, and for any point $p\in P$, let $S_p = S\cap U_p$ be
the sample points on its unit sphere.

Choose an arbitrary Delaunay pair $p_i, q_j \in P$, and let $\gamma$
be a sphere concentric with $\gamma_{ij}$ but with radius smaller
by~$1$.  This sphere is tangent to $U_{p_i}$ and $U_{q_i}$ but is at
least unit distance from every other component of $\Sigma$.  Expand
$\gamma$ about its center until it hits (without loss of generality) a
point $p'\in S_{p_i}$, and then expand it about $p'$ until it hits a
point $q'\in S_{p_i}$.  The resulting sphere $\gamma$ passes through
$p'$ and $q'$ and has no points of $S$ in its interior, so $p'$ and
$q'$ are joined by an edge in the Delaunay triangulation of $S$.
There are at least $\Omega(k^2) = \Omega(n^2\e^4)$ such edges.
\end{proof}

To create a \emph{connected} surface where good sample has a 
complicated Delaunay triangulation, we add `teeth' to our earlier 
balloon and wedge construction.  Unfortunately, in the process, we 
lose a logarithmic factor in the Delaunay complexity.

\begin{theorem}
\label{Th:teeth}
For any $n$ and any $\e < \sqrt{(\log n)/n}$, there is a smooth
connected surface~$\Sigma$ such that the Delaunay triangulation of any
parsimonious $\e$-sample of $\Sigma$ has complexity $\Omega(n^2\e^4 /
\log^2(n\e^2))$, where $n$ is the size of the sample.
\end{theorem}

\begin{proof}
Intuitively, we create the surface $\Sigma$ by pushing two rows of
regularly spaced unit balls into a large spherical balloon, similarly
to the proof of Theorem~\ref{Th:wedgie}.  As before, the surface
contains two wedges, but now each wedge has a row of small conical
teeth.  Our construction relies on the same parameters $R,w,h$ of our
earlier construction.  We now have additional parameter~$t$, which is
simultaneously the height of the teeth, the distance between the
teeth, and half the thickness of the `blade' of the wedge.

Our construction starts with the (toothless) surface described in the
proof of Theorem \ref{Th:wedgie}, but using a ball of radius $t$
instead of a unit ball to define the wedges.  We add $w/t$
evenly-spaced teeth along the blade of each wedge, where each tooth is
the Minkowski sum of a unit ball with a right circular cone of
radius~$t$.  Each tooth is tangent to both planar facets of its wedge.
To create the final smooth surface $\Sigma$, we roll a ball of radius
$t/3$ over the blade/tooth intersection curves.  The complete surface
has sample measure $\Theta((w/t)(1 + \log t) + {\log h + 1})$.
Finally, we set the parameters $w = t^2$, $h = t^3$, and $R = 20 t^3$,
so that $\mu(\Sigma) = \Theta(t\log t)$.

Let $S$ be a parsimonious $\e$-sample of $\Sigma$, and let $n =
\abs{S} = \Theta((t\log t)/\e^2)$.  For any pair of teeth, one on each
wedge, there is a sphere tangent to the ends of the teeth that has
distance $\Omega(1)$ from the rest of the surface.  We can expand this
sphere so that it passes through one point on each tooth and excludes
the rest of the points.  Thus, the Delaunay triangulation of $S$ has
complexity $\Omega(t^2) = \Omega(n^2\e^4/\log^2(n\e^2))$.
\end{proof}

\subsection{Randomness Doesn't Help Much}

Golin and Na recently proved that if $S$ is a random set of $n$ points
on the surface of a convex polytope, then the expected complexity of
the Delaunay triangulation of $S$ is $O(n)$ \cite{gn-ac3dv-00}.
Unfortunately, this result does not extend to nonconvex objects, even
the random distribution of the points is proportional to the sample
measure.

\begin{theorem}
For any $n$, there is a smooth connected surface~$\Sigma$, such that
the Delaunay triangulation of $n$ independent uniformly-distributed
random points in $\Sigma$ has complexity $\Theta(n^2/\log^2 n)$ with
high probability.
\end{theorem}

\begin{proof}
Consider the surface $\Sigma$ consisting of $\Theta(n/\log n)$ unit
balls evenly spaced along two skew line segments, exactly as in the
proof of Theorem \ref{L:balls}, with thin cylinders joining them into
a single connected surface.  With high probability, a random sample of
$n$ points contains at least one point on each ball, on the side
facing the opposite segment.  Thus, with high probability, there is at
least one Delaunay edge between any ball on one segment and any ball
on the other segment.
\end{proof}

\begin{theorem}
For any $n$, there is a smooth connected surface~$\Sigma$, such that
the Delaunay triangulation of $n$ independent random points in
$\Sigma$, distributed proportionally to the sample measure, has
complexity $\Theta(n^2/\log^4 n)$ with high probability.
\end{theorem}

\begin{proof}
Let $\Sigma$ be the surface used to prove Theorem~\ref{Th:teeth}, but
with $\Theta(n/\log^2 n)$ teeth.  With high probability, a weighted
random sample of $\Sigma$ contains at least one point at the tip of
each tooth.
\end{proof}

\section{Conclusions}
\label{S:outro}

\noindent
We have derived new upper and lower bounds on the complexity of
Delaunay triangulations under two different geometric constraints:
point sets with sublinear spread and good samples of smooth surfaces.
Our results imply that with very strong restrictions on the inputs,
existing surface reconstruction algorithms are inefficient in the
worst case.

Our results suggest several open problems, the most obvious of which
is to tighten the spread-based bounds.  Even the special case $\Delta
= \Theta(n^{1/3})$ is open.

Another natural open problem is to generalize our analysis to higher
dimensions.  Using the proof techniques in Section~\ref{SS:upper}, we
can show that any $d$-dimensional Delaunay triangulation has
$O(\Delta^{d+1})$ edges.  We conjecture that the total complexity is
always~$O(\Delta^d)$ and can only reach the maximum
$\Omega(n^{\ceil{d/2}})$ when $\Delta =\Omega(n)$.

Our bad surface examples are admittedly contrived, since they have
areas of very high curvature relative to their diameter.  An
interesting open problem is whether there are bad surfaces with
smaller `spread', \ie, ratio between diameter and minimum local
feature size.  What is the worst-case complexity of the Delaunay
triangulation of good surface as a function of the spread and sample
measure of the surface?

Our results imply that any Delaunay-based surface reconstruction
algorithm can be forced to take super-linear time, even for very
natural surface data.  It may be possible to improve these algorithms
by adding a small number of Steiner points in a preprocessing phase to
reduce the complexity of the Delaunay triangulation.  In most of our
bad surface examples, a \emph{single} Steiner point reduces the
Delaunay complexity to $O(n)$.  Bern, Eppstein, and Gilbert
\cite{beg-pgmg-94} show that any Delaunay triangulation can be reduced
to $O(n)$ complexity in $O(n\log n)$ time by adding $O(n)$ Steiner
points; see also \cite{ceghss-shcp-94}.  Unfortunately, the Steiner
points they choose (the vertices of an octtree) may make
reconstruction impossible.  In order to be usable, any new Steiner
points must either lie very close to or very far from the surface, and
as our bad examples demonstrate, both types of Steiner points may be
necessary.  Boissonnat and Cazals~(personal communication) report that
adding a small subset of the original Voronoi vertices as Steiner
points can significantly reduce the complexity of the resulting
Voronoi diagram with only minimal changes to the smooth surface
constructed by their algorithm \cite{bc-ssrnn-00}.

Very recently, Dey \etal~\cite{dfr-sralt-01} developed a surface
reconstruction algorithm that does not construct the entire Delaunay
triangulation.  Their algorithm runs in $O(n\log n)$ time if (loosely
speaking) the density of the sample points varies smoothly over the
surface.

Finally, are there other natural geometric conditions under which the
Delaunay triangulation provably has small complexity?

\paragraph{Acknowledgments.}
I thank Herbert Edelsbrunner for asking the (still open!)\ question
that started this work, Kim Whittlesey for suggesting charging
Delaunay features to area, and Edgar Ramos for suggesting
well-separated pair decompositions and sending me a copy of his
paper~\cite{dfr-sralt-01}.  Thanks also to Sariel Har-Peled, Olivier
Devillers, and Jean-Daniel Boissonnat for helpful discussions.

\def\burl#1{$\langle$\url{#1}$\rangle$}

\end{document}